\documentclass[pra,twocolumn]{revtex4}
\usepackage{bm,graphicx,amsmath,amssymb}
\usepackage{epsf,graphics,graphicx}
\usepackage{float}
\usepackage{subfigure}
\usepackage{amsthm}
\usepackage{mathrsfs}
\newcommand{\ket}[1]{|#1\rangle}
\newcommand{\bra}[1]{\langle #1|}
 
\newtheorem*{definition}{Definition}
\begin{document}
\title{Parallelity of mixed quantum ensembles}
\author{Erik Sj\"{o}qvist}
\email{erik.sjoqvist@physics.uu.se}
\affiliation{Department of Physics and Astronomy, Uppsala University, Box 516,
Se-751 20 Uppsala, Sweden}
\date{\today}
\begin{abstract}
A unifying framework for identifying distance and holonomy for decompositions of density 
operators is introduced. Parallelity between quantum ensembles is defined by minimizing 
this distance over allowed  decompositions. The minimum is a property of a pair of states 
and coincides with the Bures distance. The parallelity condition imposes a connection 
(rule for parallel transport) that results in the Uhlmann holonomy for sequences of density 
operators. A distance and holonomy for spectral decompositions of density operators is 
identified as a sub-group restriction of the full decomposition freedom. These spectral 
concepts are gauge invariant (decomposition independent) properties of mixed quantum 
ensembles, as long as the corresponding density operators are non-degenerate. A gauge 
invariant spectral geometric phase for discrete sequences of mixed quantum states is 
obtained as the phase of the trace of the spectral holonomy. This geometric phase differs 
from the interferometric mixed state geometric phase in the continuous limit. 
\end{abstract}
\maketitle

\section{Introduction}
In quantum information, quantum states act as resources for information processing, 
storage, and communication. This makes it important to quantify,  by means of, e.g.,  
coherence and correlations, the ability to perform such tasks. Geometry plays a central 
role in these quantification procedures \cite{bengtsson06}. This insight has triggered a 
revived interest in the geometry of quantum mechanical state spaces. 

Mixed quantum states may describe situations where  each particle of an ensemble in a 
quantum experiment is in a pure, but to us unknown state. A mixed quantum state can 
under these circumstances be viewed as a probability distribution $\{ r_l \}$ of pure quantum 
states $\{ \ket{\psi_l} \}$. This distribution defines a decomposition, which can be viewed as 
a set $\mathcal{A}_{\rho} = \left\{ \sqrt{r_l} \ket{\psi_l} \right\}$ of sub-normalized vectors. 
The measurable properties pertaining to the ensemble are fully described by the density 
operator $\rho = \sum_l r_l \ket{\psi_l} \bra{\psi_l}$ acting on the Hilbert space $\mathcal{H}$ 
of the system. The relation between $\mathcal{A}_{\rho}$ and $\rho$ is given by the 
projective map   
\begin{eqnarray}
\Pi : \mathcal{A}_{\rho} \mapsto \rho . 
\label{eq:projective}
\end{eqnarray}
over the set of quantum states. The map entails that $\mathcal{A}_{\rho}$ uniquely defines 
the state of the ensemble, while the converse is not true: there are infinitely many ways to 
prepare a given $\rho$. This defines a gauge structure, where gauge invariant quantities 
correspond to physical properties of mixed quantum ensembles.  

Here, we examine geometrical aspects of the projective structure in Eq.~(\ref{eq:projective}).
Specifically, we focus on the concept of {\it parallelity} associated with the projective map, and 
the resulting gauge invariant concepts of distance and holonomy along sequences of 
density operators. 

Conceptually, the present approach is based on decompositions rather than purifications 
of density operators. While the purification-based framework views a quantum ensemble as 
a part of a larger system in a pure state, no extension of the system is needed in the 
present decomposition-based scheme. This difference has been summarized in terms 
of the concepts `improper' and `proper' mixed states, as introduced by d'Espagnat 
\cite{despagnat76} (see also Ref.~\cite{timpson05}). In this sense, the present approach 
can be regarded as a proper-mixed-state framework to quantum geometry, while a 
traditional approach,  such as \cite{uhlmann86,jozsa94,nielsen00,andersson18}, 
takes the ensemble as an improper mixture of quantum states. A pertinent question in this 
context is whether the parallelity concepts associated with these `proper' (decomposition-based) 
and `improper' (purification-based) approaches mutually agree.   

The paper is organized as follows. In the next section, parallelity of decompositions 
is introduced and the resulting distance and holonomy are derived. Their relations 
to the Bures distance and Uhlmann holonomy are delineated. Section \ref{sec:spectral} 
examines the geometry when the projective map is restricted to spectral decompositions 
of non-degenerate density operators. We introduce distance, holonomy, and geometric phase 
concepts associated with such restricted projective maps. The paper ends with concluding 
remarks in Sec.~\ref{sec:conclusions}.  

\section{Parallel decompositions}
Let us start by considering two quantum ensembles, one prepared by producing $K$
orthonormal pure states $\{ \ket{e_k} \}$ with probabilities $\{ p_k \}$, and the other 
prepared by producing $L$ (not necessarily orthogonal) normalized states $\{ \ket{\psi_l} \}$ 
with probabilities $\{ r_l \}$. These two ensembles are mixtures that correspond to the same 
quantum state if and only if there exists a unitary $L \times L$ matrix $\mathbb{V}$ such that 
\cite{schrodinger36,jaynes57,hughston93} 
\begin{eqnarray}
\sqrt{r_l} \ket{\psi_l} = \sum_k \sqrt{p_k} \ket{e_k} \mathbb{V}_{kl}.
\label{eq:decomposition}
\end{eqnarray}
The sets $\{ \sqrt{r_l} \ket{\psi_l} \}$ satisfying Eq.~(\ref{eq:decomposition}) saturate the possible 
decompositions of the density operator $\rho = \sum_k p_k \ket{e_k} \bra{e_k}$ representing 
the quantum state. The decompositions are empirically equivalent as they produce the same 
set of expectation values and probabilities for all possible measurements on the ensemble.  

In general, one may have $L>K$. In such cases, only a $K \times L$ sub-matrix of 
$\mathbb{V}$ is used, while the remaining $L-K$ rows are arbitrary up to the condition that 
$\mathbb{V}$ should be unitary. In order to simplify the analysis, we shall in the following 
assume that $L=K$. In other words, it is assumed throughout that the number of states in 
the decompositions always coincides with the rank of the density operator. In order to simplify 
further, we shall assume that we compare density operators having the same rank \cite{remark1}. 

The following definition captures a notion of distance between pairs of decompositions: 
\begin{definition} 
Let $\mathcal{A}_{\rho} = \left\{ \sqrt{r_l} \ket{\psi_l} \right\}$ and $\mathcal{A}_{\sigma} = 
\left\{ \sqrt{s_l} \ket{\phi_l} \right\}$ be decompositions of two rank-$K$ density operators 
$\rho$ and $\sigma$. The distance between this pair of decompositions is defined as 
\begin{eqnarray}
D(\mathcal{A}_{\rho} , \tilde{\mathcal{A}}_{\sigma}) = 
\left( \sum_l \left|\left| \sqrt{s_l} \ket{\phi_l} -  \sqrt{r_l} \ket{\psi_l} 
\right|\right|^2 \right)^{\frac{1}{2}} .  
\label{eq:distance}
\end{eqnarray}
\end{definition} 

Based on this distance measure, we now introduce the concept of parallelity. To understand 
the idea, we note that the relevant geometrical structure is a fibre bundle for which 
decompositions form the total space, the density operators form the base space, the 
projection is the map $\Pi$ in Eq.~(\ref{eq:projective}), and the fibre is the set of unitary 
$K\times K$ matrices. 

The following definition of parallelity captures the horizontal lift of this  bundle: 
\begin{definition} 
The decompositions 
$\mathcal{A}_{\rho}$ and $\mathcal{A}_{\sigma}$ are said to be {\it parallel} if they satisfy 
\begin{eqnarray}
\sqrt{r_l} \ket{\psi_l} & = & \sum_k \sqrt{p_k} \ket{e_k} \mathbb{V}_{kl} , 
\nonumber \\ 
\sqrt{s_l} \ket{\phi_l} & = & \sum_{k,m} \sqrt{q_k} \ket{f_k} \mathbb{W}_{km} \mathbb{V}_{ml} , 
\label{eq:parallel_decompositions}
\end{eqnarray}
$\mathbb{W}$ being the unitary part of the overlap matrix 
\begin{eqnarray}
\mathbb{M}_{kl} = \sqrt{q_k} \langle f_k \ket{e_l} \sqrt{p_l} 
\end{eqnarray}
of the sub-normalized eigenvectors $\sqrt{p_l} \ket{e_l}$ and $\sqrt{q_k} \ket{f_k}$ of 
$\rho$ and $\sigma$, respectively. 
\end{definition}

We use the left polar form $\mathbb{M} = \big| \mathbb{M} \big| \mathbb{W}$ in the definition 
of $\big| \mathbb{M} \big| \geq 0$. Provided $\big| \mathbb{M} \big|^{-1}$ exists, which is 
equivalent to the stronger condition $\big| \mathbb{M} \big| > 0$, $\mathbb{W}$ is uniquely 
given by $\big| \mathbb{M} \big|^{-1} \mathbb{M}$ \cite{remark2}.  

In the following two subsections, we address the question whether the above parallelity concept 
agrees with the traditional purification-based approach. The latter is known to lead to the Bures 
distance and Uhlmann holonomy for sequences of density operators. In order to establish the 
relation between the decomposition- and purification-based approaches, we thus check explicitly 
how our parallelity concept relates to the Bures distance and Uhlmann holonomy. 

\subsection{Relation to Bures distance}
The meaning of the parallelity condition in Eq.~(\ref{eq:parallel_decompositions}) is to 
minimize the distance between decompositions in Eq.~(\ref{eq:distance}). By inserting 
Eq.~(\ref{eq:parallel_decompositions}) into Eq.~(\ref{eq:distance}), one obtains the minimal 
distance \cite{remark3}
\begin{eqnarray}
D_{\min} = \min_{\mathcal{A}_{\rho},\mathcal{A}_{\sigma}} 
D(\mathcal{A}_{\rho} , \mathcal{A}_{\sigma}) = 
\left( 2 - 2 {\rm Tr} \big| \mathbb{M} \big| \right)^{\frac{1}{2}} . 
\label{eq:minimaldistance}
\end{eqnarray}
This can be put on a familiar form by using the orthonormality of $\ket{f_k}$. We have 
\begin{eqnarray}
\sqrt{\sigma} \rho \sqrt{\sigma} & = & 
\sum_{k,l} \ket{f_k} \left| \mathbb{M} \right|_{kl}^2  \bra{f_l} 
\nonumber \\ 
 & = & \left( \sum_{k,m} \ket{f_k} \left| \mathbb{M} \right|_{km}  \bra{f_m} \right) 
\nonumber \\  
 & & \times \left( \sum_{n,l} \ket{f_n} \left| \mathbb{M} \right|_{nl}   \bra{f_l} \right) , 
\label{eq:squared_fidelity}
\end{eqnarray}
from which follows 
\begin{eqnarray}
\Big( \sqrt{\sigma} \rho \sqrt{\sigma} \Big)^{\frac{1}{2}} = 
\sum_{k,l} \ket{f_k} \left| \mathbb{M} \right|_{kl}  \bra{f_l} . 
 \label{eq:operator_fidelity}
\end{eqnarray}
We thus obtain the fidelity \cite{jozsa94}
\begin{eqnarray}
{\rm Tr} \Big( \sqrt{\sigma} \rho \sqrt{\sigma} \Big)^{\frac{1}{2}} = 
\sum_{k,l} \delta_{kl} \left| \mathbb{M} \right|_{kl} = 
{\rm Tr} \left| \mathbb{M} \right| . 
\label{eq:fidelity}
\end{eqnarray}
Equations (\ref{eq:minimaldistance}) and (\ref{eq:fidelity}) yield  
\begin{eqnarray}
D_{\min} = 
\left[ 2 - 2 {\rm Tr} \Big( \sqrt{\sigma} \rho \sqrt{\sigma} \Big)^{\frac{1}{2}} \right]^{\frac{1}{2}} , 
\label{eq:uhlmannbures}
\end{eqnarray}
which is the Bures distance $d_{{\rm B}} (\rho,\sigma)$ \cite{bures69,uhlmann76,hubner92}.  
This proves that $D_{\min}$ and the purification-based distance coincide.  

\subsection{Relation to the Uhlmann holonomy}
The parallelity condition in Eq.~(\ref{eq:parallel_decompositions}) defines a {\it connection} 
(rule for parallel transport) over the set of quantum states. It gives rise to a holonomy when 
applied to decompositions $\{ \sqrt{p_{1;l}} \ket{\psi_{1;l}} \} , \ldots , \{ \sqrt{p_{n;l}} 
\ket{\psi_{n;l}} \}$ of an ordered sequence $\mathscr{S}$ of rank-$K$ density operators $\rho_1, 
\ldots ,\rho_n$. We further assume the overlap matrices $\mathbb{M}^{(a+1,a)}$ for nearby 
decompositions satisfy $\big| \mathbb{M}^{(a+1,a)} \big| > 0$, which guarantee that all 
$\mathbb{W}^{(a+1,a)}$ are uniquely defined. The corresponding unitary $\mathbb{V}_{a}$ 
(cf. Eq.~(\ref{eq:decomposition})) is the `phase' of the decomposition. This phase can be 
transported in a parallel fashion by requiring that all nearby $\mathbb{V}_{a}$ and 
$\mathbb{V}_{a+1}$ are related according to the connection, i.e., 
\begin{eqnarray}
\mathbb{V}_{a+1} = \mathbb{W}^{(a+1,a)} \mathbb{V}_a . 
\end{eqnarray}
Iterating over the whole sequence uniquely determines the Pancharatnam-like `relative phase' 
\cite{aberg07} between the decompositions of the final and first density operators 
\begin{eqnarray}
\mathbb{V}_n \mathbb{V}_1^{\dagger} = \mathbb{W}^{(n,n-1)} \cdots \mathbb{W}^{(2,1)} . 
\label{eq:iterative_phase}
\end{eqnarray}
We can put this relative phase on more familiar form by introducing orthonormal 
eigenvectors $\ket{e_{a;k}}$ of $\rho_a$ and by observing that 
\begin{eqnarray}
\sqrt{\rho_{a+1}} \sqrt{\rho_{a}} & = & 
\left( \sum_{k,m} \ket{e_{a+1;k}} \left| \mathbb{M}_{kl}^{(a+1,a)} \right|_{km}  
\bra{e_{a+1;m}} \right) 
\nonumber \\ 
 & & \times \left( \sum_{n,l} \ket{e_{a+1;n}} \mathbb{W}_{nl}^{(a+1,a)}  \bra{e_{a;l}} \right) 
\nonumber \\ 
 & = & \Big( \sqrt{\rho_{a+1}} \rho_a \sqrt{\rho_{a+1}} \Big)^{\frac{1}{2}} 
\nonumber \\  
 & & \times \left( \sum_{n,l}  \ket{e_{a+1;n}} \mathbb{W}_{nl}^{(a+1,a)} \bra{e_{a;l}} \right) , 
\label{eq:relative_spectral_phase} 
\end{eqnarray}
where we have used Eq.~(\ref{eq:operator_fidelity}) in the last equality. By using  
Eqs.~(\ref{eq:iterative_phase}) and (\ref{eq:relative_spectral_phase}), we can express 
the Uhlmann holonomy \cite{uhlmann86} as 
\begin{eqnarray}
U_{\rm Uhl} & = & \Big( \sqrt{\rho_n} \rho_{n-1} \sqrt{\rho_n} \Big)^{-\frac{1}{2}} 
\sqrt{\rho_n} \sqrt{\rho_{n-1}} \cdots 
\nonumber \\ 
 & & \times \Big( \sqrt{\rho_2} \rho_1 \sqrt{\rho_2} \Big)^{-\frac{1}{2}} 
\sqrt{\rho_2} \sqrt{\rho_1}
\nonumber \\ 
 & = & \sum_{k_1,k_n} \ket{e_{n;k_n}} 
 \left( \mathbb{W}^{(n,n-1)} \cdots \mathbb{W}^{(2,1)} \right)_{k_1 k_n} \bra{e_{1;k_1}} 
\nonumber \\ 
 & = & \sum_{k_1,k_n} \ket{e_{n;k_n}} \left( \mathbb{V}_n 
 \mathbb{V}_1^{\dagger} \right)_{k_nk_1}  \bra{e_{1;k_1}} . 
\end{eqnarray}
Thus, our notion of parallelity in Eq.~(\ref{eq:parallel_decompositions}) is able to realize 
the Uhlmann holonomy for sequences of quantum states, again proving that the proper 
and improper approaches mutually agree. 

\section{Parallelity of spectral decompositions}
\label{sec:spectral}
As can be seen in Eq.~(\ref{eq:decomposition}), the equivalence of decompositions is 
tested relative the spectral decomposition. In this sense, the spectral decomposition 
plays a special role for mixed ensembles. This motivates us to consider the above 
distance and holonomy concepts when we restrict to the freedom left in the spectral 
decompositions of density operators.This is basically the freedom of permuting  
the eigenstates and the phase of the corresponding eigenvectors. In other words, we 
look at phases of the form 
\begin{eqnarray}
\mathbb{Y}_{kl} = e^{i\tilde{\theta}_k} \tilde{\mathbb{Q}}_{kl} ,  
\label{eq:permutationphases}
\end{eqnarray}
$\tilde{\mathbb{Q}}_{kl}$ being permutation matrices of $K$ elements. These matrices 
represent the symmetric group $S_K$. Note that $\{ \mathbb{Y} \}$ form a sub-group of 
U($K$). We shall call the distance and holonomy emerging from $\mathbb{V}$ 
the {\it spectral distance} and {\it spectral holonomy}, respectively.  

\subsection{Spectral parallelity and distance}
Assume $\left\{ \ket{e_k} \right\}$ and $\left\{ \ket{f_k} \right\}$ are two orthonormal 
sets of $K$ unit vectors and consider two non-degenerate rank-$K$ density operators 
$\rho = \sum_k p_k \ket{e_k} \bra{e_k}$ and $\sigma = \sum_k q_k \ket{f_k} \bra{f_k}$. 
Given a spectral decomposition $\mathcal{B}_{\rho} = \left\{ \sqrt{r_l} \ket{\psi_l} = 
\sqrt{p_l} \ket{e_l} e^{i\theta_l} \right\}$ of $\rho$, we look for the parallel spectral decomposition 
$\mathcal{B}_{\sigma} = \left\{ \sqrt{s_l} \ket{\phi_l} = \sum_k \sqrt{q_k} \ket{f_k} 
e^{i\tilde{\theta}_k} \tilde{\mathbb{Q}}_{kl} \right\}$ of $\sigma$. 

By varying over the permutations and phases of the spectral decomposition of $\sigma$, 
one finds the minimal distance 
\begin{eqnarray}
d_{\min} = 
\left( 2 - 2 \sum_{k,l} \sqrt{q_k p_l} \left| \langle f_k \ket{e_l} \right| \mathbb{Q}_{kl} 
\right)^{\frac{1}{2}} , 
\label{eq:minrestrdistance}
\end{eqnarray}
where the permutation $\mathbb{Q}$ satisfies 
\begin{eqnarray}
\sum_{k,l} \sqrt{q_k p_l} \left| \langle f_k \ket{e_l} \right| 
\tilde{\mathbb{Q}}_{kl} \leq \sum_{k,l} \sqrt{q_k p_l} \left| \langle f_k \ket{e_l} \right| 
\mathbb{Q}_{kl}  
\label{eq:optimalQ} 
\end{eqnarray}
for all $\tilde{\mathbb{Q}} \in S_K$. In this way, parallelity of the spectral decompositions 
can be defined as follows: 
\begin{definition} 
For non-degenerate $\rho$ and $\sigma$ \cite{remark4}, $\mathcal{B}_{\rho}$ and 
$\mathcal{B}_{\sigma}$ are said to be parallel if they satisfy 
\begin{eqnarray}
\ket{\psi_l} & = & \ket{e_l} e^{i\theta_l} , 
\nonumber \\ 
\ket{\phi_l} & = & \sum_k \ket{f_k} \mathbb{Q}_{kl} e^{i \theta_l + i\arg \langle f_k \ket{e_l}}  
\label{eq:parallel_permutation}
\end{eqnarray}
with $\mathbb{Q}$ given by Eq.~(\ref{eq:optimalQ}). 
\end{definition} 
Note that $d_{\min} \geq D_{\min}$ with equality for commuting density operators 
\cite{remark5} or pure ensembles ($K=1$). $d_{\min}$ is the spectral distance between 
$\rho$ and $\sigma$.  

\subsection{Spectral holonomy}
The parallelity of spectral decomposition defines the connection 
\begin{eqnarray}
\mathbb{V}_{a+1} = \mathbb{Q}_{kl}^{(a+1,a)} 
e^{i \arg \langle e_{a+1;k} \ket{e_{a;l}}} \mathbb{V}_a , 
\label{eq:spectral_connection}
\end{eqnarray}
being a discrete version of the connection underlying the interferometric mixed state 
geometric phase (GP) proposed in Refs.~\cite{sjoqvist00,tong04}. Equation 
(\ref{eq:spectral_connection}) describes a non-Abelian $S_K \times U(1)^{\oplus K}$ 
bundle over the space of density matrices. It yields the holonomy  
\begin{eqnarray}
U & = &  \sum_{k_1,k_n} \ket{e_{n;k_n}} 
\left( \mathbb{V}_n \mathbb{V}_1^{\dagger} \right)_{k_n k_1} \bra{e_{1;k_1}} 
\nonumber \\ 
 & = & \sum_{k_1,k_n} \ket{e_{n;k_n}} \mathbb{Q}_{k_n k_1}^{\rm tot} \bra{e_{1;k_1}} 
\nonumber \\ 
 & & \times e^{i\arg \langle e_{n;k_n} \ket{e_{n-1;k_{n-1}}} \cdots 
\langle e_{2;k_2} \ket{e_{1;k_1}}}  
\label{eq:permutation_holonomy}
\end{eqnarray} 
over the sequence $\mathscr{S}$ of order density operators $\rho_1, \ldots , \rho_n$. 
Here, $\mathbb{Q}_{k_n,k_1}^{\rm tot} \equiv \sum_{k_2,\ldots,k_{n-1}} 
\mathbb{Q}_{k_nk_{n-1}}^{(n,n-1)} \cdots \mathbb{Q}_{k_2 k_1}^{(2,1)}$ is a permutation 
matrix and we have used the identity $\arg (z_1z_2 \cdots z_n) = \arg z_1 + \arg z_2 + \ldots + 
\arg z_n$ valid for any sequence of non-zero complex numbers $z_1,z_2,\ldots,z_n$. 

To demonstrate the gauge invariance of $U$, we take the trace of the right-hand side of 
Eq.~(\ref{eq:permutation_holonomy}) over the Hilbert space and obtain 
\begin{eqnarray}
{\rm Tr} U & = & \sum_{k_1,k_n} \mathbb{Q}_{k_n k_1}^{\rm tot} 
\left| \langle e_{1;k_1} \ket{e_{n;k_n}} \right| 
\nonumber \\ 
 & & \times e^{i \arg \Delta^{(n)} (e_{1;k_1}, \ldots , e_{n;k_n})} , 
\label{eq:permutation_matrix}
\end{eqnarray} 
where each phase factor in the sum contains the phase of an $n$-point Bargmann invariant: 
\begin{eqnarray}
\Delta^{(n)} \left( e_{1;k_1}, \ldots , e_{n;k_n} \right) & \equiv &  
\langle e_{1;k_1} \ket{e_{n;k_n}} \langle e_{n;k_n} \ket{e_{n-1;k_{n-1}}} \times 
\nonumber \\ 
 & & \cdots \times \langle e_{2;k_2} \ket{e_{1;k_1}} .  
\end{eqnarray}
$\Delta^{(n)} \left( e_{1;k_1}, \ldots , e_{n;k_n} \right)$ is gauge invariant as it is 
unchanged under the transformations $\ket{e_{a;k_a}} \mapsto e^{i\chi_a} \ket{e_{a;k_a}}$, 
$a=1,\ldots,n$.  We define the {\it spectral mixed state GP} $\Phi_g$ as 
\begin{eqnarray}
e^{i \Phi_g} = \frac{{\rm Tr} U}{ \left| {\rm Tr} U \right|} , 
\label{eq:msgp}
\end{eqnarray}
for the sequence $\mathcal{S}$ of non-degenerate density operators. 

It should be noted that $e^{i \Phi_g}$ in the continuous limit differs from the interferometric 
mixed state GP \cite{sjoqvist00,tong04} that explicitly contains the spectral weights at the 
end-points as it is defined as the relative phase betwwen the internal states in the two 
interferometer beams. Thus, although being based on the same type of connection as 
the standard interferometric GP, $\Phi_g$ in Eq.~(\ref{eq:msgp}) is a different form of GP 
for mixed quantum ensembles.  

\subsection{Qubit example} 
We illustrate the spectral distance and holonomy in the case of non-degenerate isospectral 
qubit density operators $\rho_a$ with spectral decompositions $\mathcal{B}_a = \{ \sqrt{p} 
\ket{e_{a;0}} , \sqrt{1-p} \ket{e_{a;1}} \}$, $p \in [0,1]$, $p\neq \frac{1}{2}$, and $\langle e_{a;k} 
\ket{e_{a;l}} = \delta_{kl}$. The symmetric group $S_2$ consists of two elements: 
\begin{eqnarray}
\mathbb{I} = \left( \begin{array}{cc} 
1 & 0 \\ 
0 & 1 
\end{array} \right) , \ \ \ \ 
\mathbb{X} = \left( \begin{array}{cc} 
0 & 1 \\ 
1 & 0 
\end{array} \right) .   
\end{eqnarray}
The symmetry relation 
\begin{eqnarray}
\left| \langle e_{a';k} \ket{e_{a;l}} \right| = \left| \langle e_{a';l} \ket{e_{a;k}} \right|
\label{eq:visibilitysym}
\end{eqnarray}
holds for any pair $a,a'$ of qubit ensembles. 

\begin{figure}[htb]
\centering
\includegraphics[width=0.42\textwidth]{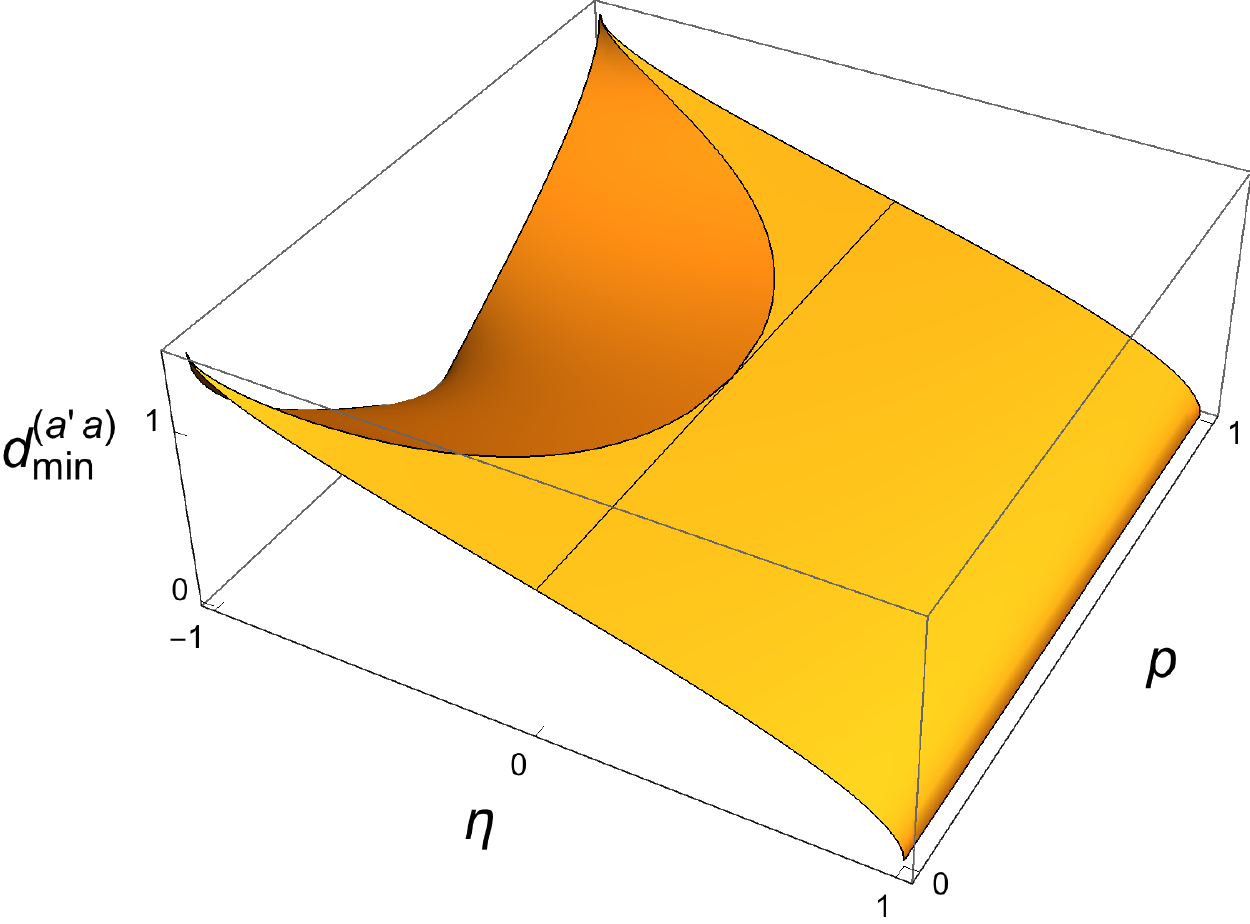}
\caption{Minimal distance $d_{\min}^{(a',a)}$ as a function of spectral weight $p$ and 
the `mutual unbiasedness' parameter $\eta = 2 \alpha \beta \cos \varphi$ (see 
Eq.~(\ref{eq:eigendecomp}) for definition of $\alpha,\beta,\varphi$). One should note that 
only $\eta = 1$ is allowed for $p=\frac{1}{2}$ since the two density operators coincide, i.e., 
we may take $\alpha \beta = \frac{1}{2}$, $\varphi=0$, in the degenerate case.}
\label{fig:distance}
\end{figure}

Let us first consider the spectral distance between $\rho_a$ and $\rho_{a'}$. We choose 
the corresponding eigenvectors to be 
\begin{eqnarray}
\{ \ket{e_{a;0}} , \ket{e_{a;1}} \} & = & \left\{ \frac{1}{\sqrt{2}} (\ket{0} + \ket{1} , 
\frac{1}{\sqrt{2}} (\ket{0} - \ket{1} \right\} , 
\nonumber \\ 
\{ \ket{e_{a';0}} , \ket{e_{a';1}} \} & = & \left\{ \alpha \ket{0} + \beta e^{i\varphi} \ket{1}, \right. 
\nonumber \\ 
 & & \left. - \beta e^{-i\varphi} \ket{0} + \alpha \ket{1} \right\} , 
\label{eq:eigendecomp}
\end{eqnarray}
where $\alpha \in [0,1]$, $\beta = \sqrt{1-\alpha^2}$, and $\varphi \in [ 0,\pi )$. These two 
eigenbases are mutually unbiased \cite{wootters89} for $\alpha \beta = 0$ 
and for $\alpha \beta = \frac{1}{2}$, $\varphi = \frac{\pi}{2}$; they are identical for 
$\alpha \beta = \frac{1}{2}$, $\varphi = 0$. To capture this, we introduce the parameter 
\begin{eqnarray}
\eta = 2 \alpha \beta \cos \varphi , 
\end{eqnarray}
$|\eta|$ being a natural measure of `mutual unbiasedness'. By using Eqs.~(\ref{eq:optimalQ}) 
and(\ref{eq:visibilitysym}), we see that $\mathbb{Q}^{(a',a)} = \mathbb{I}$ if 
\begin{eqnarray}
\left| \langle e_{a';0} \ket{e_{a;0}} \right|^2 & >& 4p(1-p) \left| \langle e_{a';1} \ket{e_{a;0}} \right|^2  
\nonumber \\ 
 & \Downarrow & 
\nonumber \\ 
1 +  \eta & > & 4p(1-p) (1-\eta)
\label{eq:I}
\end{eqnarray}
and 
$\mathbb{Q}^{(a',a)} = \mathbb{X}$ if 
\begin{eqnarray}
\left| \langle e_{a';0} \ket{e_{a;0}} \right|^2 & < & 4p(1-p) \left| \langle e_{a';1} \ket{e_{a;0}} \right|^2 
\nonumber \\ 
 & \Downarrow & 
\nonumber \\ 
1 +  \eta & < & 4p(1-p) (1-\eta) . 
\label{eq:X}
\end{eqnarray}
Since $4p(1-p) \in [0,1)$ for non-degenerate 
($p\neq \frac{1}{2}$) density operators, Eq.~(\ref{eq:I}) holds for $\eta \geq 0$ and 
$\mathbb{Q}^{(a',a)} = \mathbb{I}$ in this case. For $\eta < 0$, we find that 
$\mathbb{Q}^{(a',a)} = \mathbb{X}$ if 
\begin{eqnarray} 
\frac{1}{2} \left( 1 - \sqrt{\frac{2|\eta|}{1 + |\eta|}} \right) < p < 
\frac{1}{2} \left( 1 + \sqrt{\frac{2|\eta|}{1 + |\eta|}} \right) 
\label{eq:interval}
\end{eqnarray} 
while $\mathbb{Q}^{(a',a)} = \mathbb{I}$ otherwise. We thus see that $\mathbb{Q}^{(a',a)} = 
\mathbb{X}$ is only possible if $\eta < 0$, i.e., if $\varphi \in (\frac{\pi}{2},\pi]$. For 
$p \neq \frac{1}{2}$, we therefore find the minimal distance 
\begin{eqnarray}
d_{\min}^{(a',a)} = 
\left( 2 - 2 \sqrt{2p(1-p)} \sqrt{1 - \eta} \right)^{\frac{1}{2}} 
\end{eqnarray}
for $p$ satisfying Eq.~(\ref{eq:interval}), and 
\begin{eqnarray}
d_{\min}^{(a',a)} = 
\left( 2 - \sqrt{2} \sqrt{1 + \eta} \right)^{\frac{1}{2}} 
\end{eqnarray}
otherwise. $d_{\min}^{(a',a)}$ is shown in Fig.~\ref{fig:distance}. The non-smooth behavior 
is due to the abrupt change of the permutation of the eigenvectors that may occur for 
negative $\eta$, as described above. Note that for $p=\frac{1}{2}$ the two density operators 
coincide, i.e., by using the technique in Ref.~\cite{singh03}, we may take $\alpha \beta = 
\frac{1}{2}$, $\varphi=0$. This in turn implies that only $\eta = 1$ is allowed. In other words, 
the spectral distance is singular for degenerate eigenvalues. 

Let us now turn to the holonomy $U$ of the sequence $\mathcal{B}_1 \rightarrow 
\ldots \rightarrow \mathcal{B}_n$, $n \geq 3$. We need the pure state GPs 
$\gamma_{k_1,\ldots,k_n} = \arg \Delta^{(n)} (e_{1;k_1}, \ldots , e_{n;k_n})$, 
$k_1,\ldots,k_n = 0,1$. These phases satisfy the symmetry relation 
\begin{eqnarray}
\gamma_{k_1 \oplus 1,\ldots,k_n \oplus 1} = - \gamma_{k_1,\ldots,k_n} 
\label{eq:gpsym}
\end{eqnarray}
with $\oplus$ addition modulo 2. This implies 
\begin{eqnarray}
{\rm Tr} U & = & 2\left| \langle e_{1;0} \ket{e_{n;l}} \right| \cos \gamma_{k_1,\ldots,k_n} , 
\end{eqnarray}
where we have used Eq.~(\ref{eq:visibilitysym}). Here, $l=0$ and $l=1$ correspond to  
$\mathbb{Q}_{k_1,\ldots,k_n}^{{\rm tot}} = \mathbb{I}$ and $\mathbb{X}$, respectively. 
The sequence $k_1,\ldots,k_n$ is determined by minimizing the distance between nearby 
states. We thus see that $\Phi_g$ can only be $0$, $\pi$, or undefined. These cases 
correspond to ${\rm Tr} U > 0$, ${\rm Tr} U < 0$, and ${\rm Tr} U = 0$, respectively, or 
equivalently to whether the relevant $|\gamma|$ is smaller than, greater than, or equal 
to $\frac{\pi}{2}$. We note that this is generic in the qubit case and essentially means that 
a non-trivial spectral GP is obtained only if $|\Omega| > \pi$, $\Omega$ 
being the enclosed solid angle of the polygon on the Bloch sphere with vertices at 
$e_{1;k_1}, \ldots e_{n;k_n}$. In higher Hilbert space dimensions, the spectral GPs  
may take any value since Eqs.~(\ref{eq:visibilitysym}) and (\ref{eq:gpsym}) no longer hold. 
  
\section{Conclusions}
\label{sec:conclusions}
The decomposition freedom of a quantum state defines a projective map from the set of 
decompositions to the corresponding density operator. For pairs of states, this projective 
structure defines a natural notion of parallelity between decompositions over the states: 
parallel decompositions minimize the distance between the corresponding sub-normalized 
vectors. 

By allowing for all possible decompositions, the parallelity condition yields the Bures 
distance and Uhlmann connection. Thus, the parallelity between decompositions 
and purifications of density operators mutually agree. 

A restriction to spectral decompositions gives rise to a distance as well as a discrete 
version of the connection on which the interferometric mixed state GP in \cite{sjoqvist00,tong04} 
is based. By iteration, the connection yields a spectral holonomy for discrete sequences of 
quantum ensembles. In this way, a spectral GP is obtained as the phase of the trace 
of the spectral holonomy. As the spectral connection treats equally all density operators 
along the sequence, including its end-points, this GP does not tend to the standard 
interferometric mixed state GP in the continuous limit, as the latter explicitly involves 
the spectral weights of the initial and final states along the path. 

We end by suggesting a possible application of the proposed spectral GP and distance. 
It has been argued by Reuter {\it et al.} \cite{reuter07} that a discretized version of 
the Berry phase can be used to examine quantum critical phenomena in interacting spin 
systems in an experimentally accessible manner. This suggests that the proposed 
spectral GP and pair-distance for sets of thermal density operators could be useful 
to study phase transitions in spin systems at finite temperature. 

\section*{Acknowledgments} 
This work was supported by the Swedish Research Council (VR) through Grant No. 
2017-03832.


\begin{thebibliography}{99}
\bibitem{remark1} Parallelity between decompositions of rank $K$ and $K' \neq K$ 
desnity operators can be defined by adding $\big| K-K' \big|$ null vectors to the smallest 
rank density operator.
\bibitem{remark2} If $\big| \mathbb{M} \big|$ has at least one zero eigenvalue, then the 
decompositions are said to be {\it partially overlapping} \cite{aberg07}. Note that 
$\big| \mathbb{M} \big|$ is strictly positive if $K = \dim \mathcal{H}$.
\bibitem{remark3} The minimum of Eq.~(\ref{eq:distance}) is obtained by noting that ${\rm Re} 
\big[ {\rm Tr} \mathbb{M} \mathbb{V} ) \big] \leq {\rm Tr} |\mathbb{M}|$ with equality if and only 
if $\mathbb{V} = |\mathbb{M}|^{-1} \mathbb{M}$ for any $K \times K$ overlap matrix $\mathbb{M}$ 
such that $|\mathbb{M}| > 0$. This minimum coincides with the distance between the two 
decompositions in Eq.~(\ref{eq:parallel_decompositions}).
\bibitem{remark4} In the case of degeneracy, there is an additional unitary freedom 
\cite{singh03} in the degenerate subspace(s) when finding parallel frames.
\bibitem{remark5} In the case of commuting density operators, one finds 
\begin{eqnarray*}
\sum_{k,l} \sqrt{q_k p_l} \left| \langle f_k \ket{e_l} \right| \mathbb{Q}_{kl} = 
\sum_k \sqrt{q_k p_k} .
\end{eqnarray*}
This is the classical fidelity, which is known (see, e.g., Ref.~\cite{nielsen00}) to coincide 
with the quantum fidelity given by Eq.~(\ref{eq:fidelity}) in this particular case.
\end{thebibliography}
\end{document}